# Beam steering and topological transformations driven by interactions between discrete vortex and fundamental solitons

Xuetao Gan,[1] Peng Zhang,[1,2] Sheng Liu,[1] Fajun Xiao,[1] and Jianlin Zhao[1,*]

[1] *Key Laboratory of Space Applied Physics and Chemistry, Ministry of Education and Shaanxi Key Laboratory of Optical Information Technology, School of Science, Northwestern Polytechnical University, Xi'an 710072, China*

[2] *Currently with NSF Nanoscale Science and Engineering Center, University of California, Berkeley, CA 94720, USA*
\* Corresponding author: jlzhao@nwpu.edu.cn

**Abstract:** We study coherent and incoherent interactions between discrete vortex and fundamental solitons in two-dimensional photonic lattices, presenting a new scheme for all-optical routings and topological transformations of vorticities. Due to the multi-lobe intensity and step-phase structure of the discrete vortex soliton, the coherent soliton-interactions allow both solitons to be steered into multiple different possible destination ports depending on the initial phase of the discrete fundamental soliton. We show that charge-flipping of phase singularities and orbital angular momentum transfer can occur during the coherent interactions between the two solitons. For incoherent interactions, by controlling the relative intensities of the two solitons, we reveal that soliton-steering can be realized by either attracting the discrete fundamental soliton to four ports or localizing the four-lobe discrete vortex soliton into a ring soliton.

## 1. Introduction

Light propagation in linear and nonlinear optical waveguide arrays has been attracting great research interest [1-2] in the last decade, which represents a host for exploring discrete physics phenomena in optics, including anomalous diffraction and refraction [3-5], Anderson localization [6], Rabi oscillation [7], Zener tunneling [8], as well as photonic topological insulator [9]. With a balance of nonlinear trapping and discrete diffraction, an optical beam launched into a single site of a waveguide array can form a discrete fundamental soliton (DFS) [10-13]. The interactions between DFSs have been demonstrated to perform the blocking, deflecting and routing of optical beams along defined paths in the waveguide-array networks [14-17]. Optical vortices associated with phase dislocations and topological characters exhibit many interesting features, such as helical phase structure, doughnut intensity profile, and orbital angular momentum (OAM), which promise interesting nonlinear evolution dynamics [18,19] and a wide range of applications [20-22]. In two-dimensional (2D) nonlinear lattices induced optically and magnetically, vortices can localize into four [23,24] or more [25-27] sites and form discrete vortex solitons (DVSs). In this paper, we study the coherent and incoherent interaction dynamics between DVS and DFS in an optically induced lattice. The multi-lobe intensity and step-phase structure of DVS enable the controllable energy and angular momentum transfer during the interactions with DFS. In comparison with the interactions of two DFSs, the interactions between DVS and DFS can navigate both solitons into multiple possible destination sites, providing a promising approach for all-optical routing.

## 2. Governing equations and soliton solutions

The 2D photonic lattice is optically induced in a photorefractive crystal with a saturable screening nonlinearity [11]. The paraxial dynamics of a slowly varying optical envelope $B(x, y, z)$ along the $z$-axis in the photonic lattice is governed by the non-dimensional nonlinear Schrödinger equation [28]

$$i\frac{\partial B}{\partial z} + \frac{1}{2}\left(\frac{\partial^2 B}{\partial x^2} + \frac{\partial^2 B}{\partial y^2}\right) - \frac{E_0}{1+I_l+I_p}B = 0. \tag{1}$$

Where, $E_0$ is the applied DC electrical field, $I_p$ and $I_l$ are the intensities of the extraordinarily polarized probe beam and ordinarily polarized lattice writing beam normalized by the dark irradiance, respectively. Figure 1(a) shows the square photonic lattice determined by $I_l=I_{l0}\cos^2(\pi x/d)\cos^2(\pi y/d)$, which can be induced by four interfering plane waves experimentally [12]. Here, $I_{l0}$ and $d$ are the peak intensity and period of the lattice, which are chosen as $I_{l0}=1$, and $d=4$ in our numerical simulations. In view of typical experimental conditions [17], here one unit of $E_0$ corresponds to 1000 V/cm, and $x(y)=1$ and $z=1$ correspond to 6.4 μm and 0.88 mm, respectively.

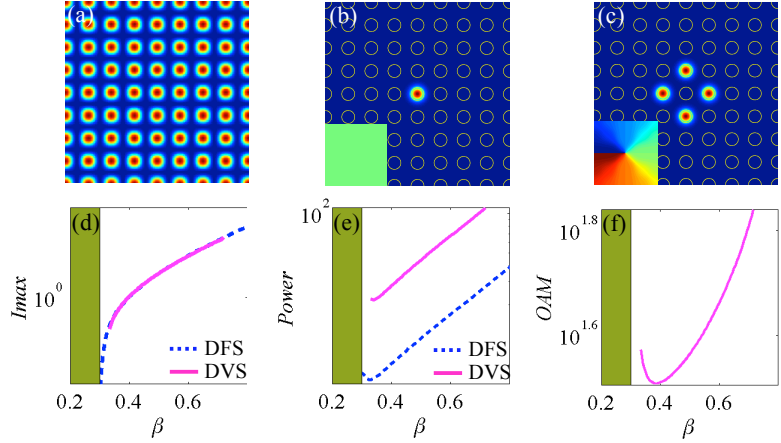

Fig. 1 (Color online) (a) Cross section of 2D square lattice; (b)-(c) intensity and phase (inset) distributions of DFS and single-charged DVS at $\beta=0.55$; (d)-(f) peak intensity, power, and OAM of solitons versus $\beta$, where dotted and solid lines correspond to the DFS and DVS respectively. Shaded area: the first Bloch band.

The solitary solutions of Eq. (1) are sought in the form of $B(x, y, z)=b(x, y)\exp(i\beta z)$, where $\beta$ is the propagation constant, and $b(x, y)$ satisfies

$$\frac{1}{2}\left(\frac{\partial^2 b}{\partial x^2} + \frac{\partial^2 b}{\partial y^2}\right) - \frac{E_0}{1+I_l+I_p}b = \beta b. \tag{2}$$

A Fourier iteration method [29] is applied to solve Eq. (2). Using a trial solution as $b(r, \theta)=r^m\exp(-r^2/36+im\theta)$, where $(r, \theta)$ are the polar coordinates and $m$ is the topological charge, we can obtain the on-site DFS and single-charged DVS with $m=0$ and 1, respectively.

Figures 1(b)-(f) display the solitons calculated under an applied electrical field of $E_0=1.5$, which provides a focusing nonlinearity for the probe beam. Both DFS and DVS are residing in the semi-infinity band gap. Different from DFSs, DVSs disappear when $\beta$ ($\beta\leq0.35$) closes to the band-edge of the first Bloch band, which indicates that DVS is not bifurcated from the Bloch waves at the band-edge [25]. Figures 1(b) and 1(c) depict the intensity and phase (insets) structures of a DFS and an on-site DVS at $\beta=0.55$. Unlike DFS, the DVS has four lobes showing $\pi/2$-step phase structure with a $2\pi$ winding phase around the central singularity. We plot the peak intensities and powers of the two solitons versus $\beta$ in Figs. 1(d) and (e). Here, the power is defined as $P=\iint|b|^2\mathrm{d}x\mathrm{d}y$. It can be seen that the peak intensities of DFS and DVS are the same monotonically increasing function of $\beta$. The power diagrams show inflexions for both solitons, implying the critical $\beta$ values for changing the soliton stability [29]. Resulting from the vorticity and singularity, the DVS possesses an OAM and its dependence on $\beta$ is shown in Fig. 1(f), where the OAM is calculated by $M=i\iint(b^*\nabla_\perp b-b\nabla_\perp b^*)\mathrm{d}x\mathrm{d}y$ [30].

The interactions of above solitons are addressed by configuring a DFS in the central site of an on-site DVS. To study the interaction dynamics, we simulate the mutual propagations of DFS ($B_1$) and DVS ($B_2$) in the photonic lattice according to Eq. (1) with the initial probe beams chosen as their exact soliton solutions [17]. Specifically, the total intensities of the probe beams are governed by $I_p=|B_1+B_2\exp(i\varphi)|^2$ and $|B_1|^2+|B_2|^2$ in the coherent and incoherent interactions, respectively, where $\varphi$ is the initial phase of DFS.

## 3. Coherent interactions

The coherent soliton interaction is sensitive to the phase difference between the solitons [31-33]. To ensure a reliable interaction process, we employ the DVS and DFS having the same propagation constant to maintain their relative phase-difference during the propagation. Figure 2 shows the interaction result of two solitons at $\beta=0.55$, where the initial phase of DFS is chosen as $\varphi=0$. Figures 2(a)-(c) depict the input soliton beams, the three-dimensional (3D) interaction trajectories in a propagation length of $z=100$, and the output beams of the DFS (1) and DVS (2), respectively. The result reveals that the coherent interaction shifts and localizes both solitons onto other ports: the DFS transfer to the port below the center lattice-site; the DVS evolves from the four-site profile to the single bottom port. Moreover, the two solitons present incessant rotations during propagating to their localized site. From the phase structures shown in the insets of Figs. 2(a) and (c), one can tell that the output DFS presents a helical phase singularity, whereas no singularity appears in the output DVS field. The rotational propagations and evolutions of phase singularity of the solitons indicate the topological transformation of vorticities during the coherent interaction.

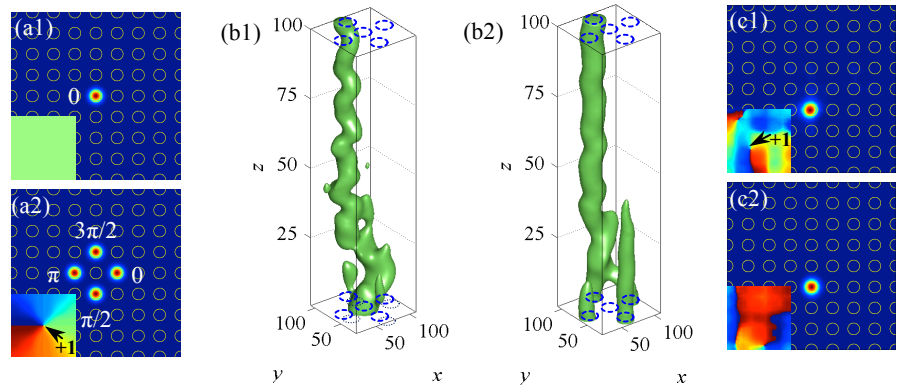

Fig. 2 (Color online) Coherent interaction of DVS and DFS at an initial phase $\varphi=0$. (a) Input solitons; (b) Iso-surface plot showing soliton interaction trajectories within the propagation length of $z=100$; (c) Output beams of the (1) DFS and (2) DVS, where the insets show the phase structures of the solitons.

To analyze the above interaction dynamics, we depict the intensity profiles of DFS (upper row) and DVS (lower row) at different propagation lengths of $z=10$, 20, and 30 in Figs. 3(a)-(c), where the insets show the phase structures. In Fig. 2(a), the phase structures of the input beams reveal that the DFS has an identical phase with the right lobe of the DVS, which constructs an attractive force between the two in-phase soliton lobes [31-33]. Moreover, the DFS is out-of-phase with the left lobe of the DVS and would move to right further due to the repulsion force [31-33]. As a result, the DFS is transferred into the adjacent right site, as shown in Fig. 3(a1). On the other hand, the attraction of the DFS breaks up the balance of energy flow of the DVS among the four lobes [34], resulting in the lobes rotate around the center lattice-site for the inherent OAM [Fig. 3(a2)]. As the original right lobe of the DVS rotates into the bottom site, the DFS is dragged into the bottom site gradually by the sustained attraction, as shown in Figs. 3(b)-(c). In further propagation, the solitons rotation is blocked by the lattice potential of the bottom site, presenting localized states for both solitons, as shown in Figs. 2(b) and (c).

An OAM transfer is expected in the coherent interaction due to the breakup of DVS and rotational propagations of the final localized solitons. We plot the calculated OAMs of DFS, DVS, and sum of them versus the propagation distance in Fig. 3(d), which are denoted as the dotted, dashed, and solid lines, respectively. It is revealed that, transferring OAM from the DVS, the DFS has a gradually increased OAM, which reaches a maximum at about $z$=20. Correspondingly, the OAM of the DVS decreases to a minimum. Next, a reverse OAM transfer process arises, *i.e.*, the DFS transfers its OAM back to the DVS as far as its value becomes zero around the propagation length of $z$=30, resulting in a maximum OAM of the DVS. The further propagation preserves the oppositely altering trends of OAM of the two solitons, implying the continuous mutual angular momentum transfer. However, the sum of OAMs of the two solitons is not conservative due to the interplay with the nonlinearity and periodicity of lattice potential [35].

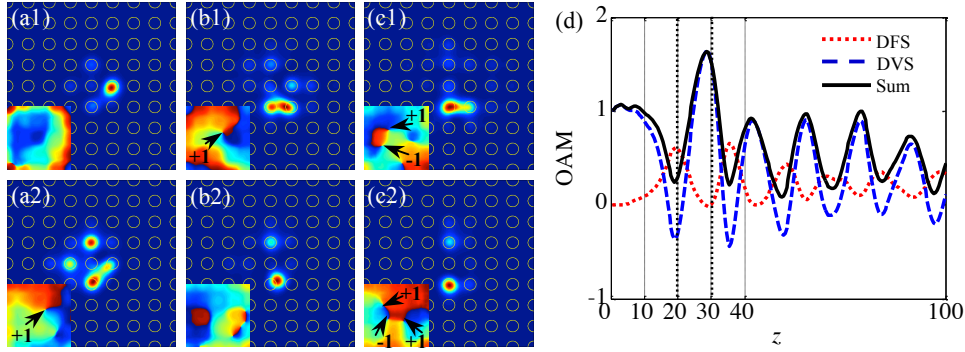

Fig. 3 (Color online) (a)-(c) Intensity and phase (insets) distributions of the DFS (upper row) and the DVS (lower row) during the coherent interaction at $z$=10, 20, and 30; (d) OAMs of DFS, DVS, and sum of them, which are marked as dotted, dashed, and solid lines, respectively.

In optical vortices, helical phase singularities with topological charge are responsible for the OAM. Following the OAM transfer, we observe the charge-flipping of phase singularities of the two solitons. We define the topological charge of the input DVS equals to +1，showing a clockwise helical phase dislocation. At the propagation distance of $z$=20, the field of DFS appears a clockwise helical phase dislocation at the center, signifying a singularity with +1 charge, as shown in Fig. 3(b1). Meanwhile, the charge of DVS becomes zero as its singularity fades away [Fig. 3(b2)]. At $z$=30, while two phase dislocations present in the DFS shown in Fig. 3(c1), their opposite rotation directions flip the total topological charge into zero. Correspondingly, the DVS has a charge of +1 after the algebraic sum of three phase dislocations with +1, -1, and +1 [Fig. 3(c2)]. Combining with Fig. 3(d), the results indicate a good agreement between the charge-flipping and OAM transfer. This topological transformation is also found at other propagation distances.

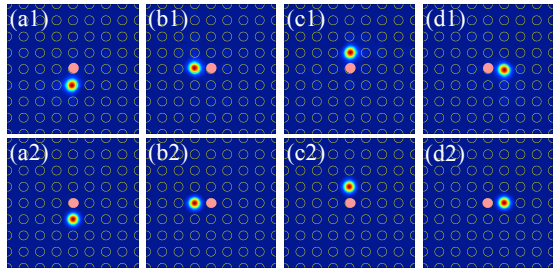

Fig. 4 (Color online) Output intensity profiles of the coherent interactions between DFS (upper row) and DVS (lower row) at $z$=100, where (a)-(d) correspond to the cases of $\varphi$=0, $\pi/2$, $\pi$, and $3\pi/2$, respectively.

Corresponding to the $\pi/2$-step phase structure of DVS, we further study its interactions with DFS having initial phases of $\varphi$=$\pi/2$, $\pi$, and $3\pi/2$, respectively. Similar evolution processes are observed under the same interaction dynamics of $\varphi$=0. Output intensity profiles of the DFS (upper row) and DVS (lower row) at $z$=100 are displayed in Fig. 4, where (a)-(d) correspond to the cases of $\varphi$=0, $\pi/2$, $\pi$, and $3\pi/2$, respectively. Note that both solitons are ultimately navigated into lower, left, upper, and right ports relative to the center lattice-site (the pink point in figures), respectively. The results demonstrate the multiple possible destination ports of the soliton-steering provided by the four lobes of DVS, and the desired route-way can be engineered by controlling the initial phase of DFS. This could be interesting for the phase-dependent optical routing and switching.

## 4. Incoherent interactions

The incoherent soliton interactions are independent of the phase difference and always yield attractive forces due to the increased refractive index in the soliton overlapping region [31], which enable effect optical blockings and routings [36]. With the benefit of the four lobes and OAM of DVS, more distinctive attraction results are expected in the incoherent interactions between DVS and DFS. We study the incoherent interactions by fixing the peak intensity of DVS as 1 and changing the peak intensity of DFS. Figure 5 shows the interaction results between DFS (1) and DVS (2), where Fig. 5(a) depicts the input soliton beams. Figures 5(b)-(c) display the 3D interaction trajectories and the output profiles of the solitons at $z$=100 for the DFS having a lower intensity of 1. The results reveal the DFS is dragged into four lobes with the same peak intensity of 0.21, indicating a possible four-port soliton-routing, where the DVS maintains its four-lobe profile. However, when the DFS intensity exceeds a threshold value (2.46 in our parameters), the attraction force from the DVS will be not strong enough to break the localization of the DFS. Choosing the peak intensity of DFS as 3.8, we obtain the interaction results shown in Figs. 5(d)-(e). Accompanying with a small broadening of the DFS, the DVS is constricted and localized onto the center lattice-site. The new DVS showing a stable ring profile is different from a ring-DVS with high intensity demonstrated in Ref. [25] and can be understood as composite solitons with the DFS [31,37]. The OAM transfer between the two solitons is not observed due to the phase-independence, however, the OAM of DVS arises a transfer from four-lobes into a single port. Numerous simulations reveal that the above behaviors qualitatively hold at other values of the DVS intensity. Compared with the coherent interaction, the independence of relative phase makes the incoherent interaction more conveniently to arrange.

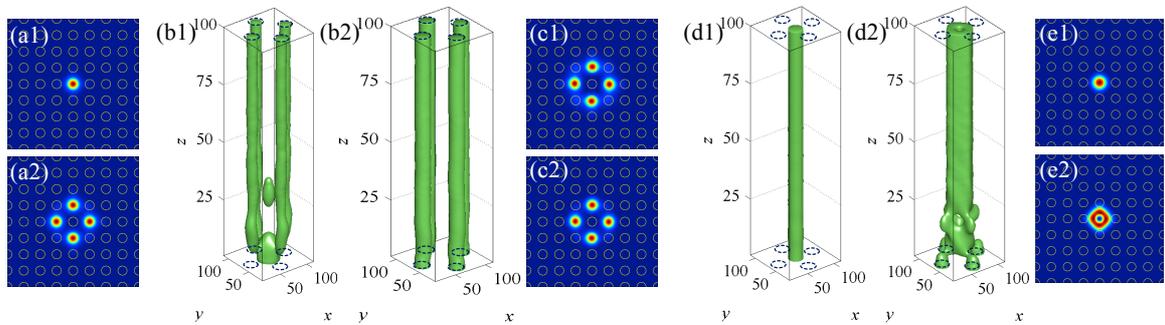

Fig. 5 (Color online) Incoherent interactions between DFS (1) and DVS (2). (a) Input solitons; (b) Iso-surface plot showing 3D soliton interaction trajectories; (c) output of the solitons at the propagation length of $z$=100, where the peak intensities of the input DFS and DVS are set to be equal; (d) and (e) are similar to (b) and (c) except the input peak intensities of the DVS and DFS are 1 and 3.8, respectively.

## 5. Conclusion

In summary, we have demonstrated soliton-steering and topological transformation behaviors during the coherent and incoherent interactions between a DVS and a DFS in 2D photonic lattices. It has been shown that the nontrivial phase and intensity structures of the DVS offer an efficient beam routing strategy with discretized destination ports. By controlling the initial DFS phase, the coherent interaction allows solitons to be routed into different desired sites. The observed charge-flipping and OAM transfer dynamics during the soliton interactions could be interesting for the OAM-based information processing in nonlinear waveguide arrays [38]. In addition, we have revealed that incoherent interactions may result in new composite solitons being either a four-lobe DFS or a ring-DVS. These rich interactions open up new prospects for all-optical routing by designing the interactions between different higher-order solitons [39], and soliton-controlling in other nonlinear periodic structures such as photonic crystals [40] and Bose-Einstein condensates [41].

**Acknowledgements**

This work was supported by the 973 Program (2012CB921900), the National Natural Science Foundations of China (61205001 and 61377035), the Natural Science Basic Research Plan in Shaanxi Province of China (2012JQ1017), the Northwestern Polytechnical University (NPU) Foundation for Fundamental Research (JC20120251), and the Technology Innovation Foundation of NPU (2011KJ01011).